\documentclass[conference,a4paper]{APSIPA2026}
\usepackage{amsmath,hyperref,flushend,amsfonts,amssymb,mathtools}
\usepackage{graphicx}
\hypersetup{hidelinks}
\usepackage[
  backend=bibtex,
  style=ieee,
  citestyle=ieee-comp,
  sorting=none
]{biblatex}
\newcommand{\R}{\mathcal{R}}
\newcommand{\Fset}{\mathcal{F}}
\newcommand{\St}{\mathcal{S}}
\newcommand{\It}{\mathcal{I}}
\newcommand{\CN}{\mathbb{C}}
\newcommand{\bx}{\mathbf{x}}
\newcommand{\ba}{\mathbf{a}}
\newcommand{\by}{\mathbf{y}}
\newcommand{\bz}{\mathbf{z}}
\newcommand{\bh}{\mathbf{h}}
\newcommand{\be}{\mathbf{e}}
\newcommand{\bp}{\mathbf{p}}
\newcommand{\bv}{\mathbf{v}}
\newcommand{\bzero}{\mathbf{0}}

\newcommand{\eps}{\varepsilon}
\DeclareMathOperator{\rank}{rank}
\DeclareMathOperator*{\argmax}{arg\,max}

\usepackage{fancyhdr}

\fancypagestyle{firststyle}{
  \fancyhf{}
  \fancyhead[C]{2026 Asia Pacific Signal and Information Processing Association Annual Summit and Conference (APSIPA ASC)}
}

\begin{document}

\title{Subspace Track-before-Detect for Passive Multi-Target Tracking with Unknown Source Signals}

\author{
\authorblockN{
Nobutaka Ito and
Yoshiaki Bando
}

\authorblockA{
National Institute of Advanced Industrial Science and Technology (AIST), Japan \\
E-mail: nobutaka.itou@aist.go.jp}
}

\maketitle
\thispagestyle{firststyle}
\pagestyle{empty}

\begin{abstract}
Passive multi-target tracking (MTT) aims to infer the time-varying kinematic and activity states of an unknown number of sources that emit unknown and possibly nonstationary signals, using only noisy mixtures of these signals observed at sensors. Track-before-detect (TBD) methods improve noise robustness by evaluating multi-target hypotheses directly on raw sensor data, without relying on a preceding detection stage. However, existing TBD likelihoods typically assume that the contribution of each active target to the observation is determined solely by its kinematic state. This assumption does not hold in passive sensing scenarios, where the observed mixtures also depend on unknown and possibly nonstationary source signals.

To address this issue, we propose \textit{subspace TBD}, a passive multi-target TBD method that employs a source-signal-insensitive likelihood derived from the complex spherical Student's $t$ (cST) distribution. Instead of explicitly modeling or estimating the nuisance source signals, the method represents each multi-target hypothesis by the subspace spanned by source steering vectors. The cST likelihood then evaluates how well the normalized multichannel mixtures align with this subspace. We conducted acoustic MTT simulations with two moving speakers in noisy, reverberant environments, comparing the proposed method with a baseline consisting of steered response power with phase transform (SRP-PHAT) followed by a sequential Monte Carlo implementation of the generalized labeled multi-Bernoulli filter (SMC-GLMB). The proposed method achieved lower mean optimal subpattern assignment (OSPA) values in all tested conditions. 
\end{abstract}

\section{Introduction}
\label{sec:intro}
This paper addresses MTT, which aims to infer the time-varying states of an unknown number of targets from noisy sensor data. These states typically include kinematic and activity states. Many MTT methods first convert sensor data into thresholded detections and then perform multi-target inference using the resulting detection sets~\cite{Reid1979,Fortmann1983,Streit1994,Mahler2003,Vo2005}. This front-end detection step is a hard and potentially lossy decision process. At low signal-to-noise ratios (SNRs), weak target evidence may be discarded by thresholding, whereas background noise or interference may be detected as spurious peaks. Because the tracker receives only the resulting detection set, sub-threshold evidence cannot be exploited in subsequent temporal filtering.

\begin{figure}[t]
  \centering
  \includegraphics[width=\linewidth]{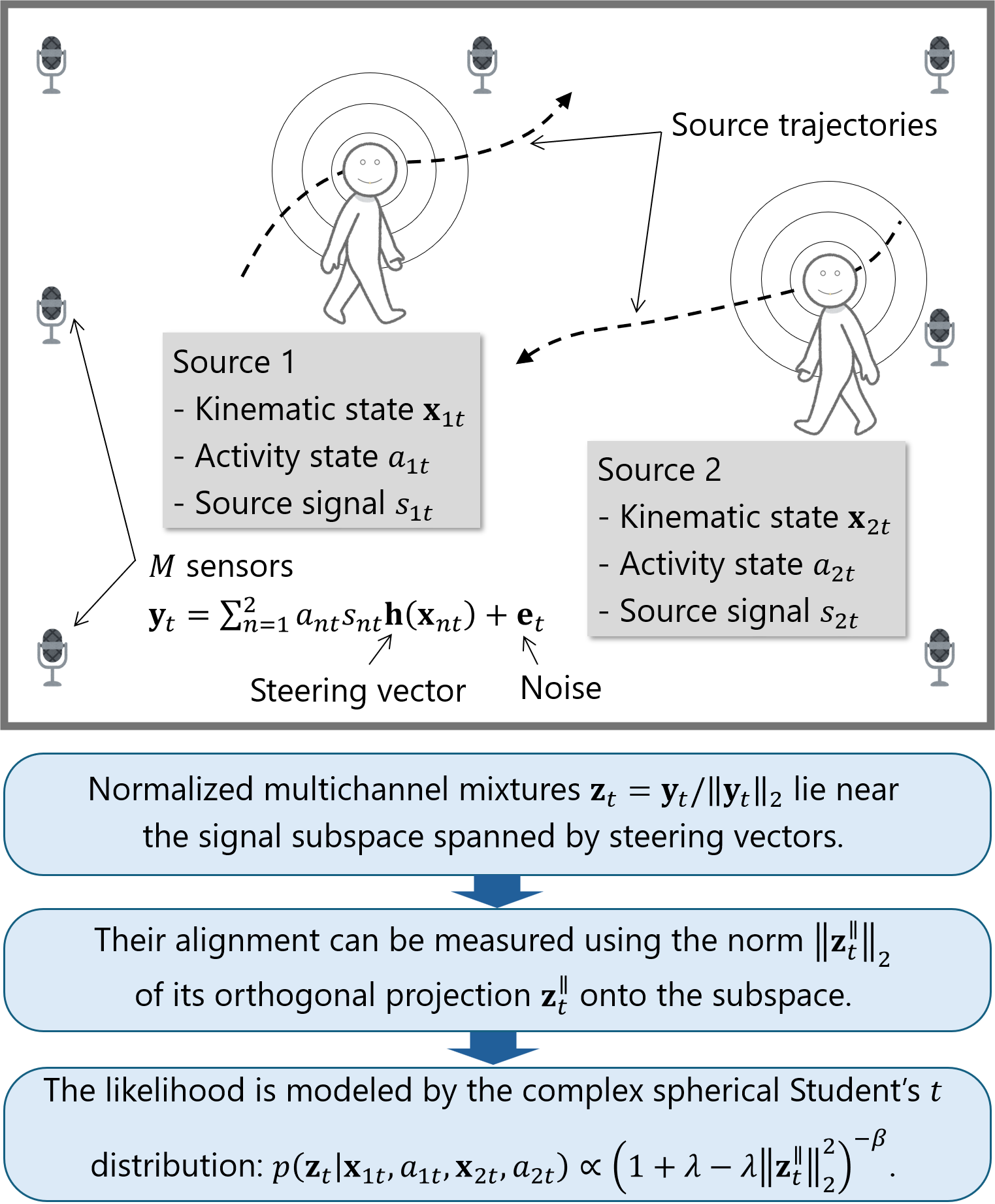}
  \caption{Passive MTT with unknown source signals and the proposed subspace likelihood, illustrated for a two-source-slot case with the frequency index omitted for clarity. Passive MTT infers the kinematic and activity states of sources from noisy multichannel mixtures whose source coefficients are unknown. After normalization, the multichannel observation direction is expected to lie near the signal subspace spanned by source steering vectors. The proposed cST likelihood evaluates each multi-target hypothesis by the squared norm of the orthogonal projection of the normalized observation onto this subspace, thereby avoiding explicit estimation of the unknown source signals.}
  \label{fig:subspaceTBD}
\end{figure}

TBD methods avoid this intermediate representation by evaluating target-state hypotheses directly from the raw sensor data and by accumulating evidence through the dynamical model~\cite{Salmond2001,Kreucher2005,Vo2010,Mahler2009,Mahler2014,Nannuru2013,Ito2020}. This makes TBD well suited to low-SNR tracking because evidence that is too weak to cross a single-frame threshold can still influence the posterior if it remains temporally consistent. However, many existing TBD formulations, especially radar and sonar TBD with known transmitted waveforms, assume that each target contribution is fully specified by its kinematic and activity states. 
Passive MTT settings, such as acoustic MTT, violate this assumption: Targets emit their own signals, such as speech, which may be unknown, nonstationary, and superposed at the sensors. In the short-time Fourier transform (STFT) domain used in this paper, the observed multichannel mixtures depend on both the targets' kinematic and activity states and the unknown STFT coefficients of the source signals. Even if the hypothesized kinematic and activity states are correct, the prediction of the sensor data can be inaccurate because the emitted signals are unknown.

Here we propose subspace TBD, a multi-target TBD particle filter (PF) designed for passive sensing scenarios involving mixed contributions of unknown source signals (Fig.~\ref{fig:subspaceTBD}). The filter represents possible sources by a fixed number of slots, each equipped with a binary activity variable governed by a birth--death Markov chain. The key observation is that the normalized multichannel sensor data lie near a low-dimensional signal subspace spanned by the steering vectors of the currently emitting sources, regardless of the nuisance source signals. We evaluate this subspace alignment using an observation likelihood based on the cST distribution~\cite{ito2026cst}, which characterizes the directional statistics of normalized complex observations. Thus, the method retains the low-SNR advantage of TBD while avoiding explicit estimation of unknown source signals. We evaluate the method in acoustic simulations with speech signals, reverberation, diffuse noise, and a competitive baseline based on SRP-PHAT~\cite{DiBiase2001} followed by the SMC-GLMB filter~\cite{Vo2013}.

\section{Proposed Subspace Track-before-Detect}
\label{sec:method}
\subsection{State and Transition Density}
The proposed filter infers source activity and kinematics jointly from microphone-array observations. Let $N$ denote the number of source slots and therefore the maximum number of simultaneously active sources represented by the filter. Slot $n$ at step $t$ has the kinematic state
\begin{equation}
\bx_{nt}= [p_{x,nt},p_{y,nt},v_{x,nt},v_{y,nt}]^{\mathsf T}\in\mathbb{R}^4
\end{equation}
and the activity variable $a_{nt}\in\{0,1\}$, where $a_{nt}=1$ indicates that the slot is active and $a_{nt}=0$ indicates that it is inactive. The stacked variables are
\begin{equation}
\bx_t=[\bx_{1t}^{\mathsf T},\ldots,\bx_{Nt}^{\mathsf T}]^{\mathsf T},\qquad
\ba_t=[a_{1t},\ldots,a_{Nt}]^{\mathsf T}.
\end{equation}
The filter infers the joint kinematic-activity state $(\bx_t,\ba_t)$ at step $t$. Its one-step transition is decomposed into an activity transition and a conditional kinematic transition,
\begin{equation}
\begin{aligned}
p(\bx_t,\ba_t\mid \bx_{t-1},\ba_{t-1})
&=\Pr(\ba_t\mid\ba_{t-1})\\
&\quad \times p(\bx_t\mid\bx_{t-1},\ba_{t-1},\ba_t).
\end{aligned}
\end{equation}
The two factors are specified below.

Activities follow independent Markov chains,
\begin{equation}
\Pr(\ba_t\mid\ba_{t-1})=
\prod_{n=1}^{N}\Pr(a_{nt}\mid a_{n,t-1}),
\end{equation}
where
\begin{equation}
\begin{cases}
\Pr(a_{nt}=1\mid a_{n,t-1}=0)=\pi_{\mathrm b},\\
\Pr(a_{nt}=0\mid a_{n,t-1}=0)=1-\pi_{\mathrm b},\\
\Pr(a_{nt}=1\mid a_{n,t-1}=1)=\pi_{\mathrm s},\\
\Pr(a_{nt}=0\mid a_{n,t-1}=1)=1-\pi_{\mathrm s}.
\end{cases}
\end{equation}
Here $\pi_{\mathrm b}$ and $\pi_{\mathrm s}$ are the birth and survival probabilities.

Given $\ba_{t-1}$ and $\ba_t$, the kinematic transition factors over slots,
\begin{equation}
p(\bx_t\mid\bx_{t-1},\ba_{t-1},\ba_t)=
\prod_{n=1}^{N}p(\bx_{nt}\mid\bx_{n,t-1},a_{n,t-1},a_{nt}).
\end{equation}
The slotwise density is
\begin{equation}
\begin{aligned}
&p(\bx_{nt}\mid\bx_{n,t-1},a_{n,t-1},a_{nt}) \\
&=\begin{cases}
\mathcal N(\bx_{nt};\mathbf F\bx_{n,t-1},\mathbf Q), & a_{n,t-1}=1,\ a_{nt}=1,\\
b(\bx_{nt}), & a_{n,t-1}=0,\ a_{nt}=1,\\
\delta(\bx_{nt}-\bx_{n,t-1}), & a_{nt}=0.
\end{cases}
\end{aligned}
\end{equation}
The matrices $\mathbf F$ and $\mathbf Q$ define the nearly constant velocity model, also known as the white-noise acceleration model, commonly used in target tracking \cite{barshalom2001tracking}. 
Newborn source states are drawn from the birth density
\begin{equation}
\label{eq:birthdensity}
b(\bx_{nt})=U_\R(\bp_{nt})\,\mathcal N(\bv_{nt};\bzero,\sigma_v^2\mathbf I_2),
\end{equation}
where $\bp_{nt}=[p_{x,nt},p_{y,nt}]^{\mathsf T}$, $\bv_{nt}=[v_{x,nt},v_{y,nt}]^{\mathsf T}$, $U_\R$ is the uniform density over the surveillance region $\R$, $\sigma_v^2$ is the newborn velocity variance, and $\mathbf I_2$ is the $2\times2$ identity matrix. The symbol $\delta$ denotes the Dirac delta function.
Inactive-slot kinematic states are dummy variables and do not enter the acoustic likelihood.

The initial state density is specified by
\begin{equation}
p(\bx_0,\ba_0)=\Pr(\ba_0)\,p(\bx_0\mid\ba_0).
\end{equation}
At $t=0$, each slot is independently active with probability $\pi_0$,
\begin{equation}
\Pr(\ba_0)=\prod_{n=1}^{N}\pi_0^{a_{n0}}(1-\pi_0)^{1-a_{n0}}.
\end{equation}
All initial kinematic states are drawn from the birth density,
\begin{equation}
p(\bx_0\mid \ba_0)=\prod_{n=1}^{N} b(\bx_{n0}).
\end{equation}

\subsection{Subspace-Based Observation and Likelihood}
Each update uses a causal block $\St_t=\{\tau_t-L+1,\ldots,\tau_t\}$ of STFT frames, where $\tau$ is the STFT-frame index and $\tau_t$ is the newest STFT frame used at step $t$. In the experiments, $L=15$. Let $\varphi_f$ be the center frequency of bin $f$, and let $\Fset$ be the set of frequency-bin indices retained in the likelihood. In the experiments, $\Fset=\{f\mid 200\,\mathrm{Hz}\le \varphi_f\le1000\,\mathrm{Hz}\}$.

Let $M$ be the number of microphones. We assume $N<M$. For $\tau\in\St_t$ and $f\in\Fset$, let $\widetilde{\by}_{\tau f}\in\CN^M$ be the vector of microphone observations in the STFT domain. Its $m$th entry is the STFT coefficient of the sound observed at microphone $m$. The narrowband signal model is
\begin{equation}
\widetilde{\by}_{\tau f}=\sum_{n=1}^{N} a_{nt}s_{n\tau f}\widetilde{\bh}_{f}(\bx_{nt})+\widetilde{\be}_{\tau f},
\end{equation}
where $s_{n\tau f}$ is an unknown source coefficient and $\widetilde{\be}_{\tau f}$ is background noise. The vector $\widetilde{\bh}_f(\bx_{nt})\in\CN^M$ is the steering vector from the source position $\bp_{nt}$ to the microphone array. Let $d_m(\bp)=\|\bp-\mathbf r_m\|_2$ and $d_{\mathrm{ref}}(\bp)=\|\bp-\mathbf r_{\mathrm{ref}}\|_2$, where $\mathbf r_m$ and $\mathbf r_{\mathrm{ref}}$ are the positions of microphone $m$ and the reference microphone, respectively. With sound speed $c$, the spherical wave model is
\begin{equation}
\begin{aligned}
\bigl[\widetilde{\bh}_f(\bx_{nt})\bigr]_m
&=\frac{d_{\mathrm{ref}}(\bp_{nt})}{d_m(\bp_{nt})}\\
&\quad\times\exp\!\left[-\mathrm j2\pi\varphi_f
\frac{d_m(\bp_{nt})-d_{\mathrm{ref}}(\bp_{nt})}{c}\right].
\end{aligned}
\end{equation}

Let $\mathbf R_f$ be a positive-definite noise coherence matrix and let $\sigma_{\tau f}^2>0$ be the noise power spectrum such that the noise covariance matrix is
\begin{equation}
\mathbb E[\widetilde{\be}_{\tau f}\widetilde{\be}_{\tau f}^{\mathsf H}]
=\sigma_{\tau f}^2\mathbf R_f.
\end{equation}
The whitening transform is
\begin{equation}
\label{eq:white_y}
\by_{\tau f}=\mathbf R_f^{-1/2}\widetilde{\by}_{\tau f},
\end{equation}
\begin{equation}
\label{eq:white_h}
\bh_f(\bx_{nt})=\mathbf R_f^{-1/2}\widetilde{\bh}_f(\bx_{nt}),
\end{equation}
\begin{equation}
\label{eq:white_e}
\be_{\tau f}=\mathbf R_f^{-1/2}\widetilde{\be}_{\tau f}.
\end{equation}
The corresponding signal model in the whitened domain is
\begin{equation}
\by_{\tau f}=\sum_{n=1}^{N}a_{nt}s_{n\tau f}\bh_f(\bx_{nt})+\be_{\tau f},
\end{equation}
\begin{equation}
\mathbb E[\be_{\tau f}\be_{\tau f}^{\mathsf H}]=\sigma_{\tau f}^2\mathbf I_M,
\end{equation}
where $\mathbf I_M$ is the $M\times M$ identity matrix.

Let $\eps>0$ be a small threshold for directional normalization and define
\begin{equation}
\It_t=\{(\tau,f)\mid \tau\in\St_t,\ f\in\Fset,
\ \|\by_{\tau f}\|_2>\eps\},
\end{equation}
\begin{equation}
\bz_{\tau f}=\frac{\by_{\tau f}}{\|\by_{\tau f}\|_2},\qquad (\tau,f)\in\It_t.
\end{equation}
Let $Z_t=\{\bz_{\tau f}\mid(\tau,f)\in\It_t\}$.

Let $K_t=\|\ba_t\|_0$ be the active cardinality. For each frequency bin $f$, the active steering matrix $\mathbf H_f(\bx_t,\ba_t)$ is the $M\times K_t$ matrix obtained by concatenating the whitened steering vectors of active slots in increasing slot order,
\begin{equation}
\mathbf H_f(\bx_t,\ba_t)=[\,\bh_f(\bx_{nt})\,]_{n:a_{nt}=1}.
\end{equation}
The orthogonal projector onto the active steering subspace is
\begin{equation}
\mathbf P_f(\bx_t,\ba_t)=\begin{cases}
\mathbf H_f(\bx_t,\ba_t)\mathbf H_f(\bx_t,\ba_t)^{\dagger}, & K_t\ge1,\\
\mathbf O_M, & K_t=0,
\end{cases}
\end{equation}
where $\mathbf O_M$ is the $M\times M$ zero matrix and $\mathbf H_f^{\dagger}$ is the Moore--Penrose pseudoinverse. Define
\begin{equation}
r_f(\bx_t,\ba_t)=\rank\{\mathbf P_f(\bx_t,\ba_t)\}.
\end{equation}
The subspace alignment statistic is
\begin{equation}
q_{\tau f}(\bx_t,\ba_t)=\bz_{\tau f}^{\mathsf H}\mathbf P_f(\bx_t,\ba_t)\bz_{\tau f},\qquad
0\le q_{\tau f}\le 1.
\end{equation}
The cST likelihood for a unit direction is
\begin{equation}
p(\bz_{\tau f}\mid\bx_t,\ba_t)=
\frac{\{1+\lambda_f[1-q_{\tau f}(\bx_t,\ba_t)]\}^{-\beta}}
{C_{M,r_f(\bx_t,\ba_t)}(\lambda_f,\nu)}.
\end{equation}
Here $\beta=(\nu+M)/2$, $\nu>0$ is the degrees of freedom, $\lambda_f=2\kappa_f/\nu$, and $\kappa_f$ is the concentration parameter. In the experiments, $\nu=2$ and
\begin{equation}
\kappa_f=A\left(\frac{\varphi_{\mathrm{ref}}}{\varphi_f}\right)^p,
\end{equation}
where $A=0.013$, $p=0.85$, and $\varphi_{\mathrm{ref}}=601.6\,\mathrm{Hz}$. The rank-dependent normalizing constant is
\begin{equation}
C_{M,r}(\lambda,\nu)=
{}_2F_1\!\left(\frac{\nu+M}{2},\,M-r;\,M;\,-\lambda\right),
\end{equation}
where ${}_2F_1$ denotes the Gauss hypergeometric function \cite{ito2026cst}. The block log likelihood is
\begin{equation}
\label{eq:blockll}
\begin{aligned}
\log p(Z_t\mid\bx_t,\ba_t)
=\sum_{(\tau,f)\in\It_t}
\Big[&-\beta\log\{1+\lambda_f[1-q_{\tau f}(\bx_t,\ba_t)]\}\\
&-\log C_{M,r_f(\bx_t,\ba_t)}(\lambda_f,\nu)\Big].
\end{aligned}
\end{equation}

\subsection{Recursive Bayesian Inference}
The filtering posterior is represented by weighted particles as
\begin{equation}
\begin{aligned}
\widehat p(\bx_t,\ba_t\mid Z_{1:t})
=\sum_{i=1}^{N_p}w_t^{(i)}\,
\delta(\bx_t-\bx_t^{(i)})
\mathbf 1\{\ba_t=\ba_t^{(i)}\}.
\end{aligned}
\end{equation}
where $N_p$ is the number of particles, $w_t^{(i)}\ge0$, and $\sum_i w_t^{(i)}=1$. At each step, candidate particles are generated by explicit proposal mechanisms. The basic proposal is the transition model in Section 1,
\begin{equation}
\begin{aligned}
&q_{\mathrm{tr}}(\bx_t,\ba_t\mid\bx_{t-1},\ba_{t-1})\\
&=\Pr(\ba_t\mid\ba_{t-1})
 p(\bx_t\mid\bx_{t-1},\ba_{t-1},\ba_t).
\end{aligned}
\end{equation}
For this proposal, the transition prior and proposal density cancel in the importance ratio. Candidates from any other proposal are weighted using the corresponding proposal correction together with the cST likelihood in \eqref{eq:blockll}. The weights are normalized and resampling is applied before the next step.

In the experiments, the transition proposal is supplemented by an SRP-PHAT proposal, where SRP-PHAT denotes steered response power with phase transform \cite{DiBiase2001}. The SRP-PHAT proposal turns selected inactive slots on, draws newborn positions from a truncated Gaussian mixture centered at selected SRP-PHAT peaks, and draws newborn velocities from the Gaussian velocity factor in \eqref{eq:birthdensity}. SRP-PHAT is used only to propose births, not as the observation likelihood.

\subsection{Point Estimation}
After the particle update, cardinality and state estimates are computed from the particle population. The cardinality posterior and slot activity probabilities are
\begin{align}
P_t(K)&=\sum_i w_t^{(i)}\mathbf 1\{\|\ba_t^{(i)}\|_0=K\},\\
\pi_{nt}&=\sum_i w_t^{(i)}a_{nt}^{(i)}.
\end{align}
The reported cardinality is $\widehat K_t=\argmax_K P_t(K)$. The $\widehat K_t$ slots with the largest $\pi_{nt}$ are declared active. For each declared active slot,
\begin{equation}
\widehat\bx_{nt}=\frac{\sum_i w_t^{(i)}a_{nt}^{(i)}\bx_{nt}^{(i)}}
{\sum_i w_t^{(i)}a_{nt}^{(i)}}.
\end{equation}

\section{Experimental Validation}
\label{sec:exp}

\subsection{Conditions}
\label{ssec:conditions}

The evaluation used a two-source acoustic scene in a
$3.0\times4.0\times2.5$\,m room. The speed of sound was
set to $343$\,m/s. Room impulse responses were simulated with
TorchRIR\footnote{\url{https://github.com/taishi-n/torchrir}}
using reverberation time $300$\,ms.
We used $M=16$ microphones placed uniformly around the rectangular
perimeter.

Speech waveforms were drawn from clean VoiceBank speech~\cite{Valentini2017}.
The selected waveforms were mean removed, variance normalized, and tiled
as needed. High-energy speech regions were retained using a $-28$ dB
threshold. The binary activity process gated whether a target emitted
at each sample. The first target was active for all $200$ tracking
updates, while the second target became active after the first $100$
updates. The tracking update interval was $\Delta t=0.128$\,s, giving
a tracking horizon of $25.6$\,s.

All mixtures were sampled at $16$ kHz and analyzed with an STFT using
a Hann window of length $1024$, hop length $512$, and FFT size $1024$.
The STFT hop corresponds to $512/16000=0.032$\,s. For each
tracking update, the proposed and conventional methods both used $15$ causal STFT snapshots ending at
that update. The SRP-PHAT front end used frequency bins from $203.125$
to $4000$ Hz, giving $244$ bins. The proposed cST likelihood used bins
from $203.125$ to $1000$ Hz, giving $52$ bins. The proposed likelihood
used a snapshot log-weight of $0.5$.

The additive noise was diffuse and spatially colored. At frequency $f$, its inter-microphone coherence was given by $[\mathbf{R}_{f}]_{mn}=\operatorname{sinc}(2\pi\varphi_fd_{mn}/c)$, where $d_{mn}$ is the microphone spacing and $\operatorname{sinc}x=\sin x/x$. A diagonal loading of $10^{-8}$ was applied before inversion. For each Monte Carlo trial, clean source image waveforms and a base diffuse-noise waveform were generated once. The noise waveform was then scaled to obtain SNRs of $10$, $0$, and $-10$ dB. Thus, for a fixed trial and SNR, the conventional baseline and the proposed method used the same microphone observations for all particle counts. The known matrix $\mathbf{R}_{f}^{-1/2}$ was used for prewhitening in both the SRP-PHAT processing and the proposed subspace likelihood.

The tracker had two target slots with a nearly constant-velocity motion model. The birth velocity standard deviation was $0.5$ m/s, and birth positions were supported within the room. The activity prior used birth probability $\pi_{\rm b}=0.02$, survival probability $\pi_{\rm s}=0.98$, and initial activity probability $0.8$. The proposed method was evaluated with $2000$, $4000$, and $8000$ particles. A weak cardinality penalty $(0,0,-0.5)$ was added for cardinalities $0$, $1$, and $2$. 

The conventional baseline applied SRP-PHAT on a $61\times61$ spatial grid, followed by an SMC-GLMB tracker~\cite{Vo2013}. Peak extraction used a relative threshold of $0.40$, at most two peaks, and a minimum peak separation of $0.32$ m. The multi-target tracker used detection probability $0.75$, clutter rate $6.0$, birth weight $0.20$, score-sum birth weighting, Gaussian localization-error standard deviation $0.28$ m, and peak-score power $1.5$. The baseline shared the dynamics, array geometry, colored-noise knowledge, particle-count settings, and evaluation protocol with the proposed method.

Ten Monte Carlo trials were used for each SNR and particle count. Randomness came from speech selection, trajectory generation, diffuse-noise generation, and particle filtering. The target-activity schedule was used only to generate data and evaluate the estimates. It was not supplied to the filters. Performance was measured by the position OSPA metric~\cite{Schuhmacher2008} with cutoff $c=1$ m and order $p=2$, averaged over time. 

\subsection{Results}
\label{ssec:results}

\begin{figure*}[t]
  \centering
  \includegraphics[width=0.7\textwidth]{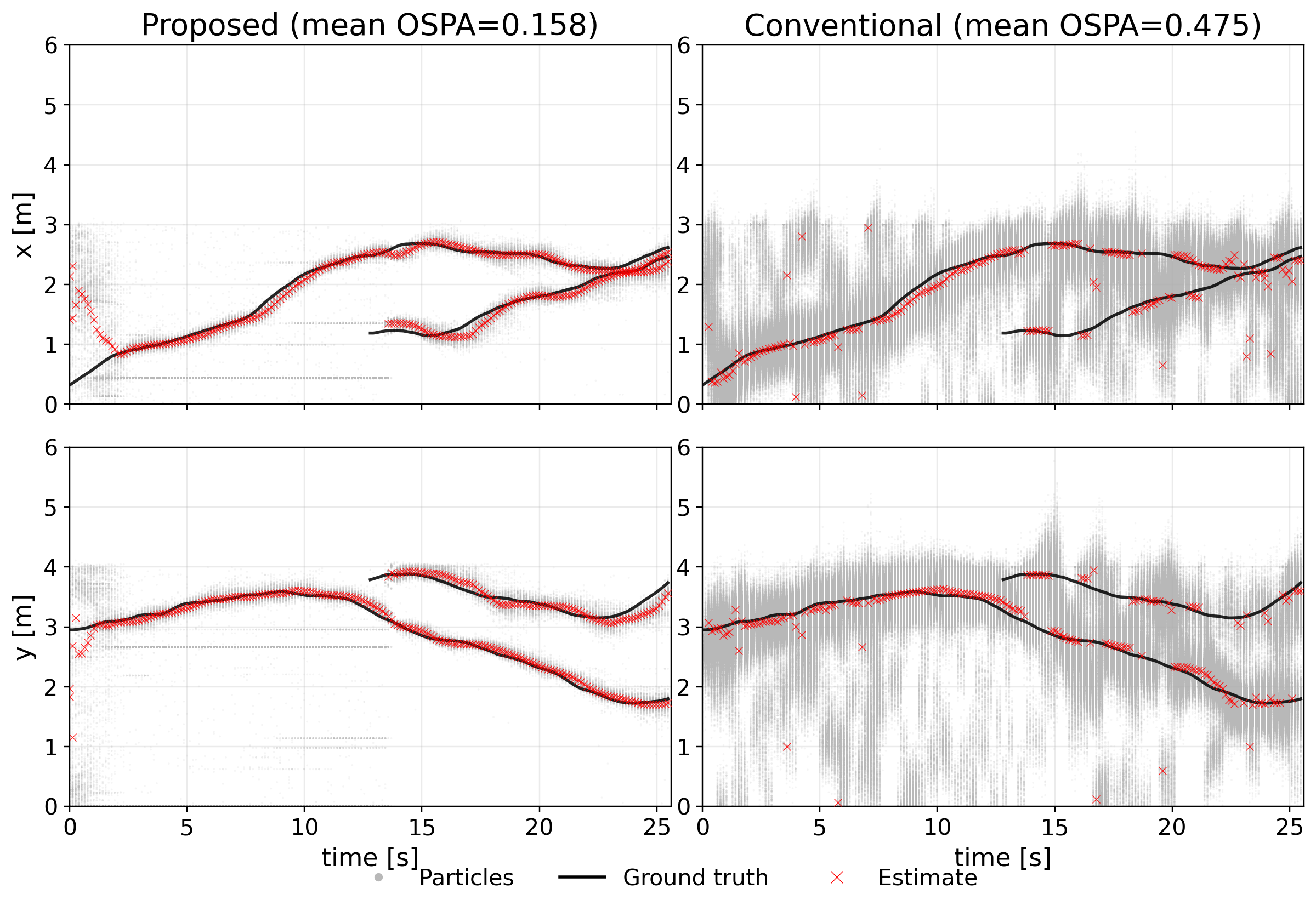}
  \caption{Representative time-position plots for a reverberant speech scene with $-10$\,dB SNR and $4000$ particles. Black curves show the ground-truth source positions, red crosses show extracted estimates, and pale gray points show subsampled particle locations. The proposed {subspace TBD} estimate follows the ground-truth coordinate histories more closely than the conventional SRP-PHAT+SMC-GLMB baseline in this trial.}
  \label{fig:representative_tracks}
\end{figure*}

Figure~\ref{fig:representative_tracks} shows a representative $-10$\,dB trial with $4000$ particles. The conventional SRP-PHAT+SMC-GLMB baseline produced a number of displaced estimates and had a mean OSPA of $0.475$ m in this sequence. The proposed filter had a mean OSPA of $0.158$ m. Its particle cloud was more concentrated around the two coordinate histories, especially after the second source became active. 

\begin{figure}[t]
\centering
\includegraphics[width=\linewidth]{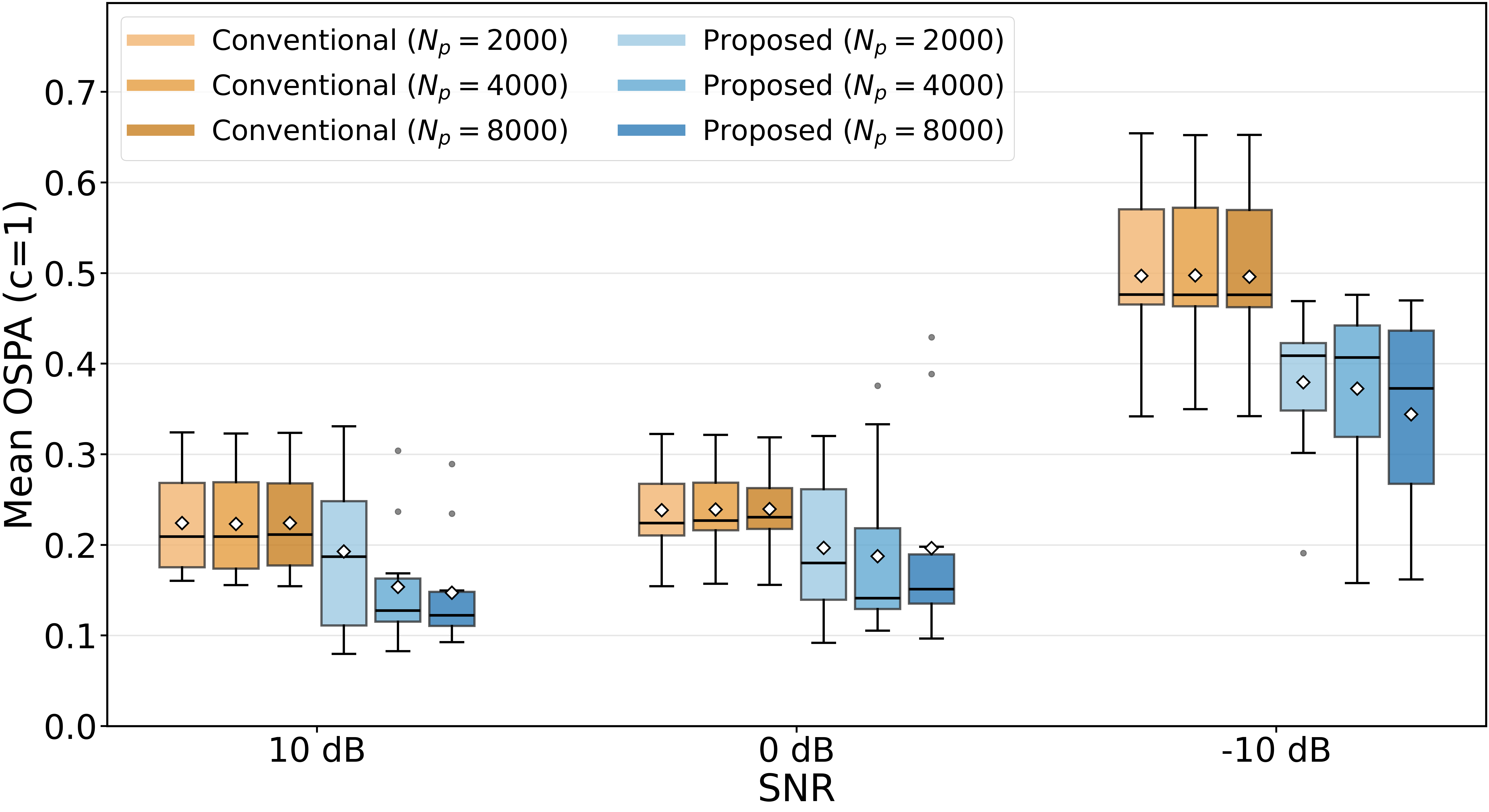}
\caption{Position-set OSPA distributions for the main reverberant speech experiment with diffuse spatially colored noise and ten Monte Carlo trials per condition. Boxes show trials, white diamonds show means, and lower values are better.}
\label{fig:main_boxplot}
\end{figure}

As in Figure~\ref{fig:main_boxplot}, the proposed method achieved lower mean OSPA than the conventional baseline in all nine tested SNR and particle-count settings. At $10$ dB, the proposed means were $0.193\pm0.091$, $0.154\pm0.068$, and $0.147\pm0.064$ m for $2000$, $4000$, and $8000$ particles, while the corresponding baseline means were $0.224\pm0.058$, $0.223\pm0.058$, and $0.224\pm0.057$ m. At $0$ dB, the proposed means were $0.197\pm0.079$, $0.188\pm0.095$, and $0.196\pm0.116$ m, compared with baseline means of $0.238\pm0.052$, $0.239\pm0.051$, and $0.240\pm0.049$ m. The largest gains occurred at $-10$ dB, where the proposed means were $0.379\pm0.085$, $0.372\pm0.103$, and $0.344\pm0.110$ m, while the baseline means were $0.497\pm0.097$, $0.497\pm0.096$, and $0.496\pm0.098$ m.

\subsection{Discussion}
\label{ssec:discussion}

The main comparison suggested that the subspace likelihood was most helpful when the localization front end was stressed by low SNR. The proposed filter evaluated particles directly against the whitened multichannel observations. This allowed weak but temporally consistent subspace evidence to affect both the kinematic posterior and the activity posterior.

The proposal mechanism should be interpreted as an efficiency device rather than as a change of observation model. SRP-PHAT peaks were useful because they concentrated birth candidates near likely source positions. However, the strict mixture-density correction was essential. Without the correction, a sharper proposal would have implicitly altered the birth-position prior and could have biased the activity estimate toward the front-end peak count. Retaining the rank-dependent cST normalizing constant addressed a related issue on the likelihood side by reducing the tendency to favor higher-cardinality hypotheses solely because they span larger subspaces.

Finally, the acoustic model mismatch was intentional. Reverberant speech mixtures were generated with room impulse responses, whereas the filter evaluated direct-path steering subspaces. The results therefore showed usefulness under controlled mismatch, not exactness of the direct-path model. The current evidence should be read as a two-source simulation study with known array geometry and known noise covariance. Measured room responses, online covariance adaptation, and more sources remain future work.

\section{Conclusion}
\label{sec:conc}

This paper presented {subspace TBD} for passive MTT with unknown source signals. The method jointly estimates activity and kinematic states by combining a birth--death activity model, PF inference, and the cST subspace likelihood with rank-dependent normalization. Reverberant speech simulations with diffuse noise showed lower mean OSPA than the SRP-PHAT+SMC-GLMB baseline in all tested settings, with the largest gains at $-10$ dB. Remaining limitations include real recordings, more than two simultaneous sources, and online adaptation of room transfer functions, noise covariance, and cST concentration parameters.

\printbibliography

\end{document}